\documentclass[12pt]{article}
\usepackage{amsmath,amssymb}
\usepackage{amsthm}
\usepackage{hyperref}
\usepackage{overpic}

\usepackage[latin1]{inputenc}

\newcommand{\id}{\hbox{1\kern-.20em l}}

\theoremstyle{remark}

\begin{document}

\begin{flushright}
 {\tt ULB-TH/11-09}
\end{flushright}

\begin{center}  {\bf \Large On the symmetry orbits of black holes in non-linear sigma models}

\vspace{.9 cm}

 Josef Lindman H\"ornlund\footnote{jlindman@ulb.ac.be}

\vspace{.9 cm}

{\em Service de Physique Th\'eorique et Math\'ematique,\\
Universit\'e Libre de Bruxelles \& International Solvay Institutes\\ Campus Plaine C.P. 231, Boulevard du
Triomphe, B-1050 Bruxelles, 
Belgium}
\end{center}

{\footnotesize

{\abstract
Breitenlohner, Maison and Gibbons claimed some time ago that all bona-fide four dimensional asymptotically flat non-degenerate black holes are in a symmetry orbit of the Schwarzschild/Kerr black hole in a large set of theories of gravity and matter. Their argument involved reducing the theory on a time-like Killing vector field and analysing the resulting three dimensional sigma model of maps to a symmetric space $G/H$. In the construction of their proof, they conjectured the existence of a suitable $H$-transformation that always remove the electromagnetic charges of the four dimensional black hole solution. We show in this short note that such a transformation does not exist in general, and discuss a set of boundary conditions on the horizon for the scalar fields in the sigma model that yield black holes for which the result by Breitenlohner, Maison and Gibbons can be applied.}
}

\section{Introduction}

Three dimensional sigma models have proven to be a powerful tool when investigating and generating black holes in four or five dimensional theories of gravity coupled to some appropriate matter content. For solutions that are invariant under the action of a time-like Killing field, a Geroch-like \cite{Geroch:1970nt} reduction to a three dimensional Euclidean orbit space is possible. In theories where the scalar fields in four dimensions parametrise a symmetric space $G_4/H_4$ the resulting three dimensional theory can be rewritten, after dualising one-forms, as a theory of gravity coupled to a non-linear sigma model of maps to a larger symmetric space $G/H$, such that $G_4/H_4 \subset G/H$. This type of theories were discussed and classified in some detail in the paper \cite{Breitenlohner:1987dg} by Breitenlohner, Maison and Gibbons. In three dimensions solutions to the sigma-model transform non-linearly under the Lie group symmetry $G$. The subgroup of $G$ that preserve asymptotic flatness is the isotropy subgroup $H$ and we can therefore use $H$ to classify solutions in terms of orbits of this subgroup. Breitenlohner {\it et.al}. used these Lie group transformations in the derivation of a theorem stating that all non-degenerate four-dimensional black holes are in the orbit of the Schwarzschild or Kerr black hole, in the static and stationary case respectively. The derivation goes as follows:
The charges of the four dimensional solution are characterised by an element $X \in \mathfrak{g}$ where $\mathfrak{g}$ is the Lie algebra of $G$ (the details are reviewed in section \ref{sec:symmetric}). The isotropy subgroup $H$ act on $X$ by conjugation. Under the assumption that a $H$ transformation exists that remove the electromagnetic charges of $X$, it was proven in \cite{Breitenlohner:1987dg} that the transformed solution must be the Schwarzschild (or the Kerr) solution. Although the result is stated as a theorem, the authors were aware of a possible hole in the argument, as stated by Maison in \cite{Maison:2000fj} for example. The main result of this short note is to show in section \ref{sec:harrison} by a rather elementary argument that the conjectured $H$ transformation does not exist in general. We need therefore to impose a suitable set of regularity conditions on the solutions in the sigma model. For static and spherically symmetric black hole solutions, we discuss in section \ref{sec:complex} a set of such conditions, and show that after imposing them, the arguments of \cite{Breitenlohner:1987dg} can be applied. We end in section \ref{sec:conclusions} by a short discussion about our result and interesting generalisations.

\section{Symmetric spaces and black holes}
\label{sec:symmetric}

Let us review the sigma model machinery and its relation to black holes. Let $G$ be a semi-simple, connected and real Lie group. An irreducible symmetric space $G/H$ is characterised by an involution $\Sigma : G \rightarrow G$ such that $F_0^{\Sigma} \subset H \subset F^{\Sigma}$ where $F^{\Sigma}$ is the subgroup of elements fixed by $\Sigma$ and $F_0^{\Sigma}$ the component of $F^{\Sigma}$ connected to the identity. The push-forward $\sigma = (\Sigma_*)_e$ induce a {\it canonical decomposition}
\begin{equation}
\mathfrak{g} = \mathfrak{h} + \mathfrak{m}
\end{equation}
such that $\mathfrak{h}$ and $\mathfrak{m}$ are the sub-vector-spaces of elements in $\mathfrak{g}$ with eigenvalues $+1$ and $-1$ under the action of $\sigma$, respectively. Hence $\mathfrak{h}$ is the Lie algebra of the Lie group $H$ and $\mathfrak{m}$ transform in an irreducible representation of $\mathfrak{h}$. From \cite{Berger:1957} we know that there exists a Cartan involution $\theta$, commuting with $\sigma$, giving rise to the local Cartan decomposition
\begin{equation}
\mathfrak{g} = \mathfrak{k}+ \mathfrak{p},
\end{equation}
where $\mathfrak{k}$ is the Lie algebra of the maximally compact subgroup $K$ of $G$ and $\mathfrak{p}$ the vector space of non-compact elements in $\mathfrak{g}$. Since $\sigma$ and $\theta$ commute, we can decompose $\mathfrak{g}$ further as 
\begin{equation}
\mathfrak{g} =  \mathfrak{h}\cap \mathfrak{k}+\mathfrak{h} \cap \mathfrak{p}+  \mathfrak{m}\cap \mathfrak{k}+ \mathfrak{m}\cap \mathfrak{p}
\end{equation}
and we define $\mathfrak{f} =\mathfrak{m} \cap \mathfrak{p}$. The space $\mathfrak{f}$ is the vector space of non-compact elements in $\mathfrak{m}$. 

Let $\mathcal{L}_4$ be the Lagrangian of a four dimensional theory in the classification of \cite{Breitenlohner:1987dg}, such that the scalars in this theory parametrise a symmetric space $G_4/H_4$. This four dimensional theory, reduced to three dimensions over the orbits of a time-like Killing vector field $\xi$, gives rise to a non-linear sigma model of maps to $G/H$, coupled to three dimensional gravity. Letting $\xi = \partial_t$ we can write the four dimensional line element as
\begin{equation}
\mathrm{d}s^2 = - 2U (\mathrm{d}t + \omega_3 )^2 + \tfrac{1}{2}U^{-1} \mathrm{d}s^2(\mathcal{M}_3),
\end{equation}
where 
\begin{equation}
\label{eqn:norm}
U =- \tfrac{1}{2} g(\xi, \xi)
\end{equation}
is the norm of $\xi$. We write the Lagrangian living on the three dimensional manifold $\mathcal{M}_3$ as
\begin{equation}
\label{eqn:3dlagrangian}
\mathcal{L}_3 = R_3 \cdot \mathrm{vol}_3+ \mathrm{tr}_F(\mathrm{d}M \wedge \star_3 \mathrm{d}M^{-1})
\end{equation}
where $M : \mathcal{M}_3 \rightarrow G$ is a symmetric map such that 
\begin{equation}
\label{eqn:symmetric}
\Sigma(M) = M^{-1}, 
\end{equation}
$\mathrm{tr_F: \mathfrak{g}}\times \mathfrak{g} \rightarrow \mathbb{R}$ is the invariant bilinear form on $\mathfrak{g}$ in the fundamental representation of $G$ acting on a vector space $V$, and $R_3$ is the three dimensional Ricci scalar. The map $M$ and the line element $\mathrm{d}s^2(\mathcal{M}_3)$ fully determine four dimensional solutions invariant under the action of $\xi$. Let $r$ be a radial co-ordinate on $\mathcal{M}_3$. If $M$ describe an asymptotically flat black hole, we have
\begin{equation}
\label{eqn:asymptoticexpansion}
M = 1 + \frac{X}{r} + \mathcal{O}\bigg(\frac{1}{r^2}\bigg)
\end{equation}
near spatial infinity. The Lie algebra element $X$ is now in $\mathfrak{m}$ since $M$ obeys (\ref{eqn:symmetric}). We can furthermore associate the entries of $X \in \mathfrak{m}$ to the four dimensional conserved charges, such as mass, NUT charge, electromagnetic charges and scalar charges. Recall that Ehlers's reduction give rise to the a $\mathrm{SL}(2, \mathbb{R})/\mathrm{SO}(2)$ sigma-model \cite{Ehlers:1957zz}. Together with $G_4/H_4$ this coset form the totally geodesic submanifold
\begin{equation}
F = \mathrm{SL}(2, \mathbb{R})/\mathrm{SO}(2) \times G_4/H_4
\end{equation}
such that $F = \exp \mathfrak{f}$. Hence $\mathfrak{f}$ parametrises the mass $m$, the NUT charge $n$ and the scalar charges. The complement $\mathfrak{m} \cap \mathfrak{k}$ parametrises the electromagnetic charges. For the solution to be non-degenerate and obey a four dimensional mass bound, we demand that
\begin{equation}
\label{eqn:positivenorm}
\mathrm{tr}_F (X^2) > 0 .
\end{equation}
For a more thorough discussion about these points, see \cite{Breitenlohner:1987dg, Bossard:2009at} for example.

The Lie subgroup $H$ act on $M$ by conjugation, and it is easy to see that conjugation leaves the Lagrangian (\ref{eqn:3dlagrangian}) invariant, and preserve the form (\ref{eqn:asymptoticexpansion}), i.e. asymptotic flatness. This implies that $H$ act non-linearly on the four dimensional solutions, and by conjugation on $X$. From the canonical decomposition of $\mathfrak{g}$ we know that conjugation preserves $\mathfrak{m}$. Let us now proceed to analyse the orbits of $H$ on $\mathfrak{m}$.

\section{A suitable Harrison-transformation does not always exist}
\label{sec:harrison}

The statement of \cite{Breitenlohner:1987dg} (and repeated in for example \cite{Bossard:2009at}) that a $H$ transformation always exists that removes the electromagnetic charges can now be restated as the claim that the $H$-orbit of elements $X \in \mathfrak{m}$ with positive norm always intersects $\mathfrak{f}$. Let us analyse this in some detail. The eigenvalues of $X$ in the fundamental representation can be calculated by finding the roots $\lambda_i$, $i = 1,...,n$ to the characteristic equation
\begin{equation}
\label{eqn:eigenvalues}
{\det}_F (X - \lambda_i \cdot 1) = 0 .
\end{equation}
The integer $n$ is the dimension of the vector space $V$. Since
\begin{eqnarray}
{\det}_F(Ad_h X - \lambda_i \cdot 1)& =& {\det}_F ( h (X - \lambda_i \cdot 1) h^{-1}) \nonumber \\
&=& {\det}_F (h) {\det}_F (X - \lambda_i \cdot 1) {\det}_F (h^{-1}) \nonumber \\
&=& {\det}_F(X - \lambda_i \cdot 1),
\end{eqnarray}
it follows that equation (\ref{eqn:eigenvalues}) is invariant under the action of $H$. Eigenvalues of $X$ are hence constant over the $H$-orbit. Every element $Y \in \mathfrak{f}$ has real eigenvalues in the fundamental representation since $\theta(Y) = - Y$. However, the condition (\ref{eqn:positivenorm}) is not sufficient to ensure that this is true for a general $X\in \mathfrak{m}$.\footnote{The exception here is the case of Einstein-Maxwell, or equivalently the $\mathrm{SU}(2,1)/(\mathrm{SL}(2, \mathbb{R}) \times \mathrm{U}(1))$ sigma-model, where the trace square is the only invariant polynomial and determine the eigenvalues of $X$ completely \cite{Simon:1985hk, Maison:1984tg}. It is likely that this singular case might have inspired an unwarranted generalisation to more involved symmetric spaces.} The eigenvalues of $X$ are determined in general by higher order invariant polynomials such as ${\det}_F (X)$ and $\mathrm{tr}_F(X^4)$ as well. There are therefore plenty of elements in $\mathfrak{m}$ with imaginary eigenvalues and positive norm, and a Harrison transformation can never remove the electromagnetic charges from these elements, or equivalently; $Ad_hX$ has necessarily some leg left in $\mathfrak{m} \cap \mathfrak{k}$ independently of how we choose $h \in H$. Consider for example pure five-dimensional gravity reduced to three dimensions, giving rise to the $\mathrm{SL}(3, \mathbb{R})/\mathrm{SO}(2,1)$ sigma-model. Here the important polynomial is ${\det}_F(X)$ and its easy to find elements $X$ with positive norm with respect to $\mathrm{tr}_F$ and non-zero determinant that have some imaginary eigenvalues in the fundamental representation. We can conclude that for the argument of \cite{Breitenlohner:1987dg} to be applicable, some regularity conditions on $X$ must be imposed. In the next section we briefly discuss such a condition.

\section{Complex eigenvalues and boundary conditions}
\label{sec:complex}

Consider static and spherically non-degenerate symmetric black hole solutions in four dimensions. We argue in this section that it is impossible to satisfy a certain set of boundary conditions at the black hole horizon if the element $X$ has complex eigenvalues in the fundamental representation. The argument is rather elementary but requires some further notation. In the case of static and spherically symmetric black holes, the solutions to the equations of motion of the Lagrangian (\ref{eqn:3dlagrangian}) become geodesics on $G/H$ (embedded in $G$) and are given by
\begin{equation}
M = \exp_G \tau  X
\end{equation}
for $X \in \mathfrak{m}$ and $\tau : \mathcal{M}_3 \rightarrow \mathbb{R}$ a certain function depending only on the radial co-ordinate $r$ on $\mathcal{M}_3$. The function $\tau$ is parametrised such that $\tau(\infty) = 0$ and $\tau(r_{\mathcal{H}}) = \infty$ if $r_{\mathcal{H}}$ is the position of the horizon. Let $\Phi$ be the set of scalars in the three dimensional sigma model and let $U \in \Phi$ be the norm (\ref{eqn:norm}) of the Killing vector field we reduce upon, with respect to the four dimensional metric $g$. From \cite{Breitenlohner:1987dg} (eq. (4.15) section 4), we know that there is a set of boundary conditions one can impose at the horizon. Under these boundary conditions, all scalars in $\Phi$ take a finite value at the black hole horizon $r_{\mathcal{H}}$. In particular $U$ vanish at the horizon, since the horizon is a Killing horizon with respect to $\xi$. This implies that we can write 
\begin{equation}
M = \Sigma(\mathcal{V}_{rest})^{-1} \exp_G( \log(U) h) \mathcal{V}_{rest}
\end{equation}
where $h \in \mathfrak{f}$ is the Cartan element in Ehlers's $\mathfrak{sl}(2, \mathbb{R})$ (up to some suitable normalisation) and $\lim_{\tau \rightarrow \infty}\mathcal{V}_{rest} = \mathcal{V}_0$.

Let $\eta : V \times V \rightarrow \mathbb{R}$ be a non-degenerate metric on $V$. As in the previous section, let $\lambda_i$ be the eigenvalues of $X$. Since the characteristic equation of $X$ only has real coefficients, all complex eigenvalues come in conjugate pairs. Let $\lambda_1$ and $\lambda_2$ be such a complex pair, and $\epsilon_1$ and $\epsilon_2$ the corresponding eigenvectors. This implies that $M$ act on the vectors $y_1 = \epsilon_1+ \epsilon_2$ and $y_2 = \sqrt{-1}(\epsilon_1 - \epsilon_2$) as
\begin{eqnarray}
M^{a} y_1 = e^{a\, Re \lambda_1\, \tau}\big(\cos(a\, Im \lambda_1\, \tau) y_1 - \sin(a\,  Im \lambda_1\, \tau)y_2\big) \nonumber \\
M^{a} y_2 = e^{a\, Re \lambda_1\, \tau}\big(\cos(a\, Im \lambda_1\, \tau) y_2 + \sin(a\, Im \lambda_1\, \tau ) y_1\big),
\end{eqnarray}
where we chose $a =1$ if $Re \lambda_1 > 0$ and $a=-1$ if $Re \lambda_1 < 0$. Since $\eta$ is non-degenerate we can always find $v \in V$ such that $\eta(v, y_1) \neq 0$. We thus have
\begin{eqnarray}
\label{eqn:limit1}
\eta(v, M^ay_1) = e^{a \, Re \lambda_1\,  \tau}\big( \cos(a\, Im \lambda_1\, \tau) \eta(v, y_1) - \sin(a\, Im \lambda_1\, \tau) \eta(v, y_2)\big)
\end{eqnarray}
and 
\begin{eqnarray}
\label{eqn:limit2}
\eta(v, M^ay_1) & = & \eta(v, \Sigma(\mathcal{V}_{rest})^{-a} \exp_G( a \log(U) h) {\mathcal{V}_{rest}}^a y_1).
\end{eqnarray}
We know that the right hand side of (\ref{eqn:limit1}) oscillate between larger and larger positive and negative values, as $\tau \rightarrow \infty$ and $Im \lambda_1 \neq 0$ by assumption. Since $\mathcal{V}_{rest}$ approach a fixed matrix $\mathcal{V}_0$ and $U \rightarrow 0$, this can never be the case for the right hand side of (\ref{eqn:limit2}). We conclude that complex eigenvalues are not compatible with the boundary conditions of \cite{Breitenlohner:1987dg}.

Let us continue by showing that a semi-simple element $X \in \mathfrak{m}$ with only real eigenvalues always is conjugate to $\mathfrak{f}$. Since $\mathfrak{g}$ can be decomposed into real eigenspaces of $X$ there is a Cartan involution $\theta'$ such that $\theta'(X) = -X$. According to Lemma 3 in \cite{Matsuki:1979} there exists a $g \in G$ such that $Ad_g X = X$ and $Ad_g \theta'$ commute with $\sigma$. From \cite{Berger:1957} we know that all Cartan involutions commuting with $\sigma$ are conjugate by $H$, and this is equivalent to $X$ being conjugate to $\mathfrak{f}$.

\section{Conclusions and discussion}
\label{sec:conclusions}

We have thus shown in section \ref{sec:harrison} that a Harrison transformation that takes us to $\mathfrak{f}$ does not always exist.  In order to classify black hole solutions in terms of a non-linear indefinite sigma-model, we must therefore impose a regularity condition on all solutions before we are able to use the results of \cite{Breitenlohner:1987dg} on the uniqueness of the $H$-orbit of the Schwarzschild solution. In section \ref{sec:complex} we showed that if we require all scalars to take a finite value at the horizon, all geodesics describing regular static and spherically symmetric black holes must have tangent vectors without complex eigenvalues. By the conjugacy of Cartan involutions we finally showed that such tangent vectors are conjugate to $\mathfrak{f}$ via $H$ and the uniqueness theorems 6.2 and 6.3 of the Schwarzschild solution in the gravity and scalar section of \cite{Breitenlohner:1987dg} can be applied. 

Some questions remain however and could be interesting to investigate further. For five dimensional cohomogeneity one black holes, the boundary conditions at the horizon might need to be generalised (see for example \cite{Hornlund:2010tr}). The argument in section \ref{sec:complex} should however be fairly easy to extend also to this case. It might also be take case that there are interesting solutions that do not obey this set of boundary conditions, or generalisations thereof. 

It is furthermore not clear how conjugation by $H$ preserves regularity in general. One can find simple examples in which $H$ change the sign of the mass of the black hole, and in this case take a regular black hole to a singular one. It is therefore possible that a geodesic on $F$ could correspond to a singular solution but conjugation of the geodesic to $G/H$ would yield a regular black hole, or vice versa.

Finally, it would be interesting to investigate how the discussion in this note apply to stationary, but not static, black holes, in an attempt to put Theorem 7.3 in \cite{Breitenlohner:1987dg} on some more solid ground.

\subsubsection*{Acknowledgements}

We are indebted to Amitabh Virmani, Axel Kleinschmidt and Guillaume Bossard for helpful comments.

\bibliographystyle{utphys}
\bibliography{Refdata}

\end{document}